\pdfoutput=1 
\documentclass{JINST}

\usepackage{amsmath}

\usepackage{subfigure}
\usepackage{booktabs}
\usepackage{lineno}

\usepackage{multirow}

\newcommand{\Neq}{\ensuremath{\text{n}_{\text{eq}}\,\text{cm}^{-2}}}
\newcommand{\irrad}[2]{\ensuremath{#1\cdot10^{#2}\,\Neq}}

\title{Measurements on HV-CMOS Active Sensors After Irradiation to HL-LHC fluences}

\author{B. Ristic$^a\,^b$ on behalf of the ATLAS CMOS pixel collaboration\\
\llap{$^a$}CERN, Geneva, Switzerland\\
\llap{$^b$}Universit\'{e} de Gen\`{e}ve, Geneva, Switzerland
E-mail: \email{branislav.ristic@cern.ch}}

\abstract{During the long shutdown (LS) 3 beginning 2022 the LHC will be upgraded for higher luminosities pushing the limits especially for the inner tracking detectors of the LHC experiments. 
In order to cope with the increased particle rate and radiation levels the ATLAS Inner Detector will be completely replaced by a purely silicon based one. 
Novel sensors based on HV-CMOS processes prove to be good candidates in terms of spatial resolution and radiation hardness. 
In this paper measurements conducted on prototypes built in the AMS H18 HV-CMOS process and irradiated to fluences of up to \irrad{2}{16} are presented.}


\begin{document}

\section{Novel semiconductor sensors for the ATLAS Inner Tracker}\label{sec:novelsemi}
The detectors of the ATLAS\footnote{A Toroial LHC ApparatuS} Inner Detector, the Pixel Detector, the Semiconductor Tracker and TRT\footnote{Transition Radiation Tracker}\cite{ATLAS} are the subdetectors closest to the interaction point.
Therefore, they have to cope with the highest density of particles and highest radiation levels.
After the Phase II upgrade the LHC\footnote{Large Hadron Collider} is expected to deliver up to $3000\,\text{fb}^{-1}$ of integrated luminosity to the ATLAS Detector leading to a total ionizing dose beyond $1\,$GRad and NIEL\footnote{NonIonizing Energy Loss} fluences beyond \irrad{2}{16} for the innermost Pixel layer \cite{atlasloi}.
At the expected particle rates the TRT will not be able to operate efficiently anymore.
Therefore the new inner tracker will be built based entirely on silicon sensors.
For the inner layers high granularity combined with an extremely radiation hard design is necessary to fulfill the set requirements while going to larger radii the cost of producing large areas becomes the limiting factor for detector construction.\\
A novel approach of producing silicon sensors emerged from exploiting High-Voltage CMOS\footnote{Complementary Metal-Oxide-Semiconductor} processes as they are commonly used in industry for high voltage switching electronics.
  \begin{figure}[htbp]
    \centering
    \includegraphics[width=.8\textwidth]{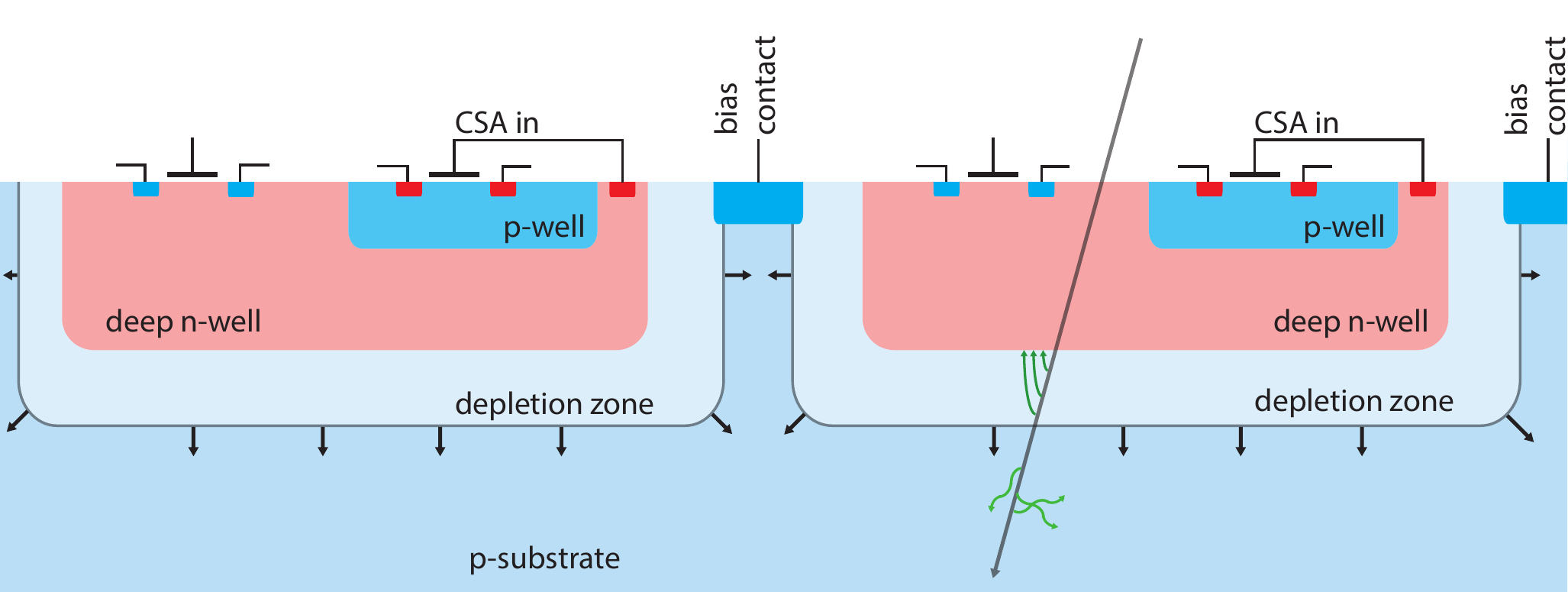}
    \caption{Simplified cross section of an HV-CMOS sensor. The highly sensitive depletion zone is growing from the deep n-well implants that shield the electronics sitting on the surface of the chip from the high voltage in the bulk\cite{phd_backhaus}}
    \label{fig:hvcmos}
  \end{figure}
Here deep n-wells are implanted in a low resistivity p-type substrate shielding the sensitive electronics from the high voltage of the bulk and acting as collecting electrodes for charge deposited in the bulk.
As the substrate is highly conductive the high voltage can be applied from the front side of the chip, eliminating the need for a back side processing, thus lowering the cost of production.
The highly doped substrate is intrinsically radiation hard, but allows for only shallow depletion zones.
Amplification and further signal processing electronics are therefore implemented inside the deep n-well.\\
Current pixel sensors are usually bump bonded to a dedicated readout chip forming a hybrid detector.
Due to the small pitch of $50\,\mu$m this process is expensive, but can be avoided by coupling the sensor capacitively to the readout chip simply by gluing.
Although capacitive coupling is in principle possible for passive sensors, the signal amplitude is usually not sufficient for the attached readout chip.
CMOS sensors allow to choose the output amplitude due to the on chip electronics, delivering a good signal to noise ratio for the readout chips.

\section{Pixel sensors in AMS 180nm High Voltage CMOS technology}
\begin{figure}[!ht]
  \begin{center}
    \subfigure[]{
      \includegraphics[width=0.47\textwidth]{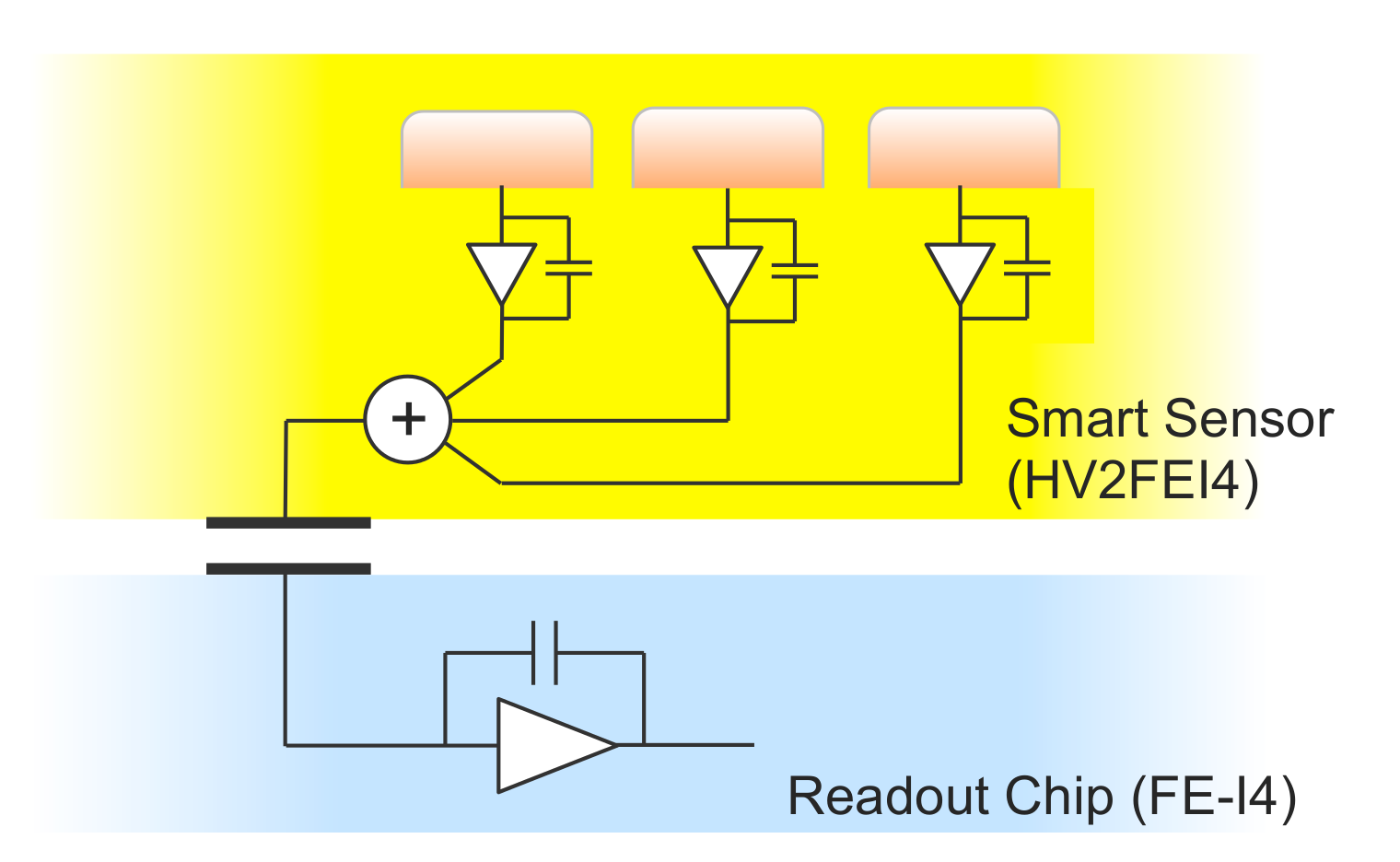}
      \label{fig:ccpd}
    }
    \subfigure[]{
      \includegraphics[width=0.42\textwidth]{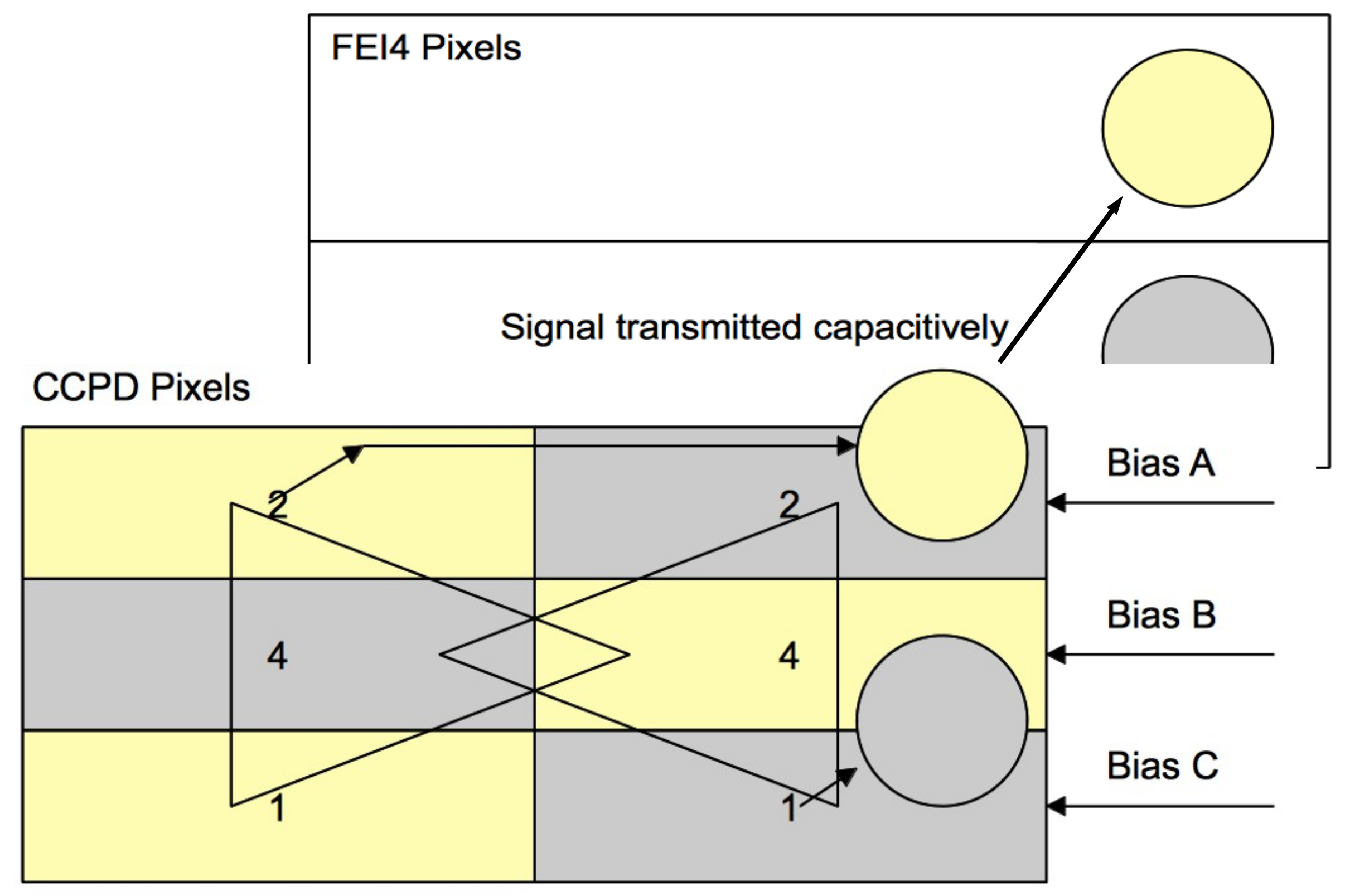}
      \label{fig:chessboard}
    }
    \vspace{-6pt}
    \caption[]{\subref{fig:ccpd} Sketch of a capacitively coupled HVCMOS sensor. Three subpixels are connected to one readout electrode that is forming a capacitor with its counterpart on the readout chip. The mapping of the subpixels to the FE-I4 readout cells is depicted in \subref{fig:chessboard}} 
  \end{center}
\end{figure}

The prototypes investigated in this paper were built in the Austria MicroSystems 180nm HV-CMOS process based on designs by I. Peric\cite{hvcmos}.
AMS offers deep n-well structures implanted in a low resistivity ($\sim 10\, \Omega \text{cm}$) p-type substrate allowing for a bias voltage of up to $-90\,$V which is expected to yield a depletion depth in the order of $15\,\mu$m.
For MIP\footnote{Minimally Ionizing Particle}-like particles this results in a deposited charge with a MPV\footnote{Most Probable Value} of around $1200\,$ electrons.
The so called HV2FEI4 sensors were adapted to the latest ATLAS pixel readout chip, the FE-I4\cite{fei4} and typically contain multiple pixel flavors implementing various optimizations for radiation hardness and low noise levels.
In the bigger part of the pixel matrix usually a structure consisting of a charge sensitive amplifier combined with a two stage discriminator is situated.
Threshold levels can be set globally as well as local corrections can be applied by setting a per pixel DAC\footnote{Digital Analog Converter}.
For the first two prototype versions that will be discussed here each group of three $125x33\,\mu$m pixel cells is connected to one readout pad in order to match the FE-I4 footprint (see figure \ref{fig:chessboard}).
Output stages of the subpixels are globally adjustable allowing an encoding of the hit position in the analogue output signal and therefore in the recorded charge, measured in TOT\footnote{Time over Threshold} for that FE-I4 pixel cell (see figure \ref{fig:ccpd}).
Version 1 was a proof of principle design deliberately using standard cells while the successor v2 was adapted to withstand elevated radiation levels.
The prototypes proved to be operational after X-Ray irradiation of up to $862\,$MRad\cite{peric2014}.
The third prototype version allows to access an $100 \times 100 \, {\mu m}^2$ deep n-well via a wirebond pad and allows for undisturbed measurements of the bulk signal like described in chapter \ref{seq:etct}.
Irradiations for the presented prototypes have been done with thermal neutrons at the TRIGA reactor of JSI Ljubljana\cite{ljubljana}.

\section{Testbeam results}\label{seq:tbresults}
Testbeam measurements have been performed with unirradiated and irradiated HV2FEI4 version 1 and 2 samples at the DESY\footnote{Deutsches ElektronenSYnchrotron}, the CERN\footnote{Conseil Europ\'een pour la Recherche Nucl\'eaire} PS\footnote{Proton Synchrotron} and SPS\footnote{Super Proton Synchrotron} testbeam lines using an EUDET JRA1 and a new FE-I4 based telescope. The measurements were concentrated on the ``standard pixels'', that compose the bigger part of the pixel matrix and aim for low noise at a reasonably high radiation tolerance.
At the time the testbeams were performed, the sub pixel decoding was not implemented into the readout, so no information about the hit sub-pixel can be extracted.
This poses a difficulty for the alignment during the reconstruction of the data and was circumvented partly by virtually merging each pair of FE-I4 pixels that correspond to a unit cell resulting in virtual pixels of $250x100\,\mu\text{m}^2$ size.
However, a frame of seemingly lower efficient pixels at the edge of the sensor remains as an artifact.
These pixels were not taken into account for analysis.
\begin{table}[htbp]
\begin{center}
\caption{List of samples measured in testbeams in 2013 and 2014}
\begin{tabular}{cccc}
\toprule
\multirow{2}{*}{Sample} & HV2FEI4 & Irradiation fluence & \multirow{2}{*}{Testbeam period}\\
& version & [\Neq] & \\
\midrule
C07 & 1 & $1\cdot10^{15}$ & DESY 2013\\
C19& 2 & $0$ & CERN PS 2014, CERN SPS 2014\\
C22 & 2 & $1\cdot10^{15}$ & CERN SPS 2014\\
\bottomrule
\end{tabular}
\label{tab:samples}
\end{center}
\end{table}

\begin{figure}[!ht]
    \begin{center}
      \subfigure[]{
	\includegraphics[width=0.47\textwidth]{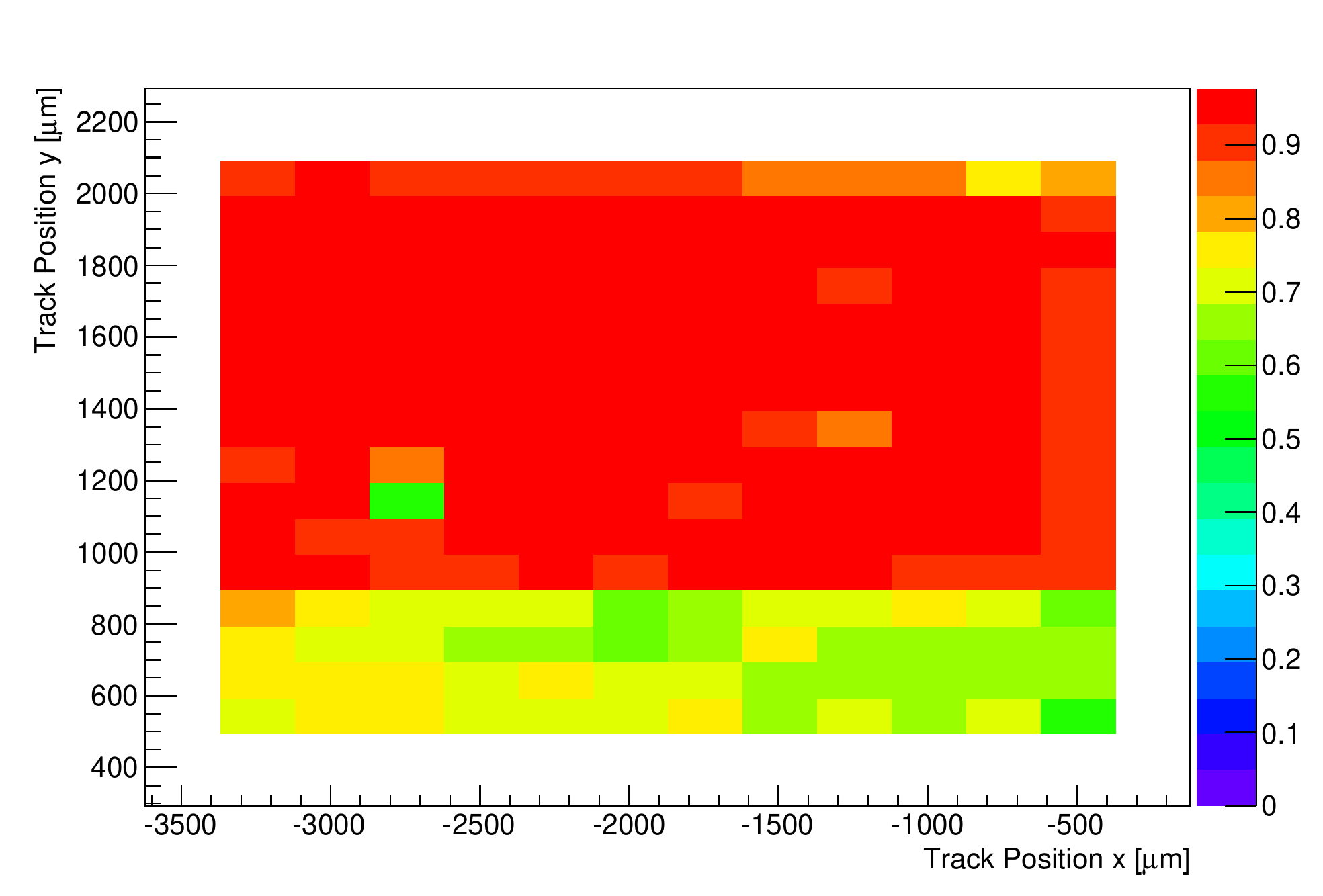}
	\label{fig:ps_c19_eff}
      }
      \subfigure[]{
	\includegraphics[width=0.48\textwidth]{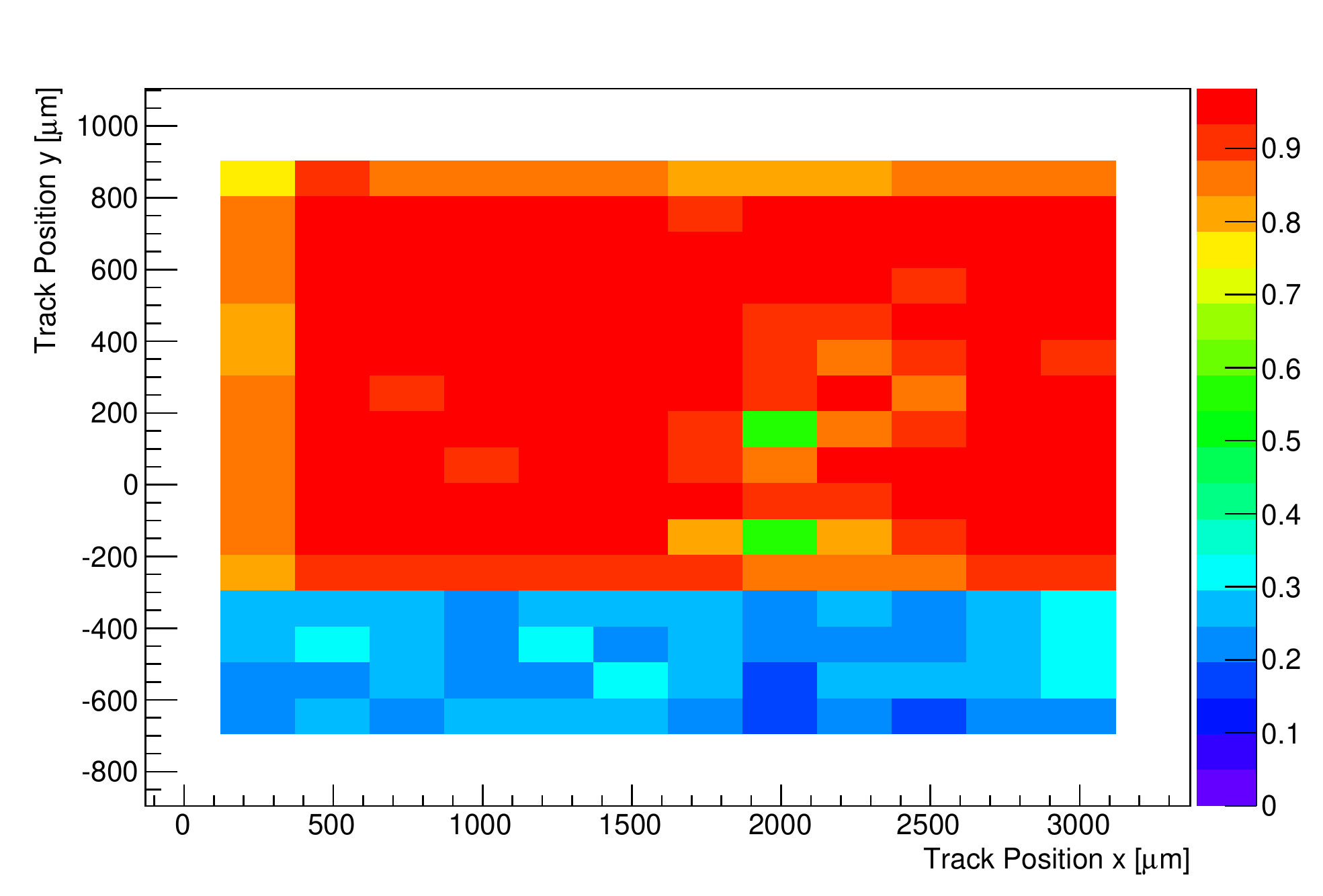}
	\label{fig:sps_c22_eff}
      }
      \vspace{-6pt}
      \caption[]{Detection efficiency maps for \subref{fig:ps_c19_eff} C19 (CERN PS testbeam) and \subref{fig:sps_c22_eff} C22 (CERN SPS testbeam). The low efficiency regions at the bottom side of the plots originate from another pixel type with a higher threshold.} 
    \end{center}
  \end{figure}

\begin{figure}[!ht]
    \begin{center}
      \subfigure[]{
	\includegraphics[width=0.48\textwidth]{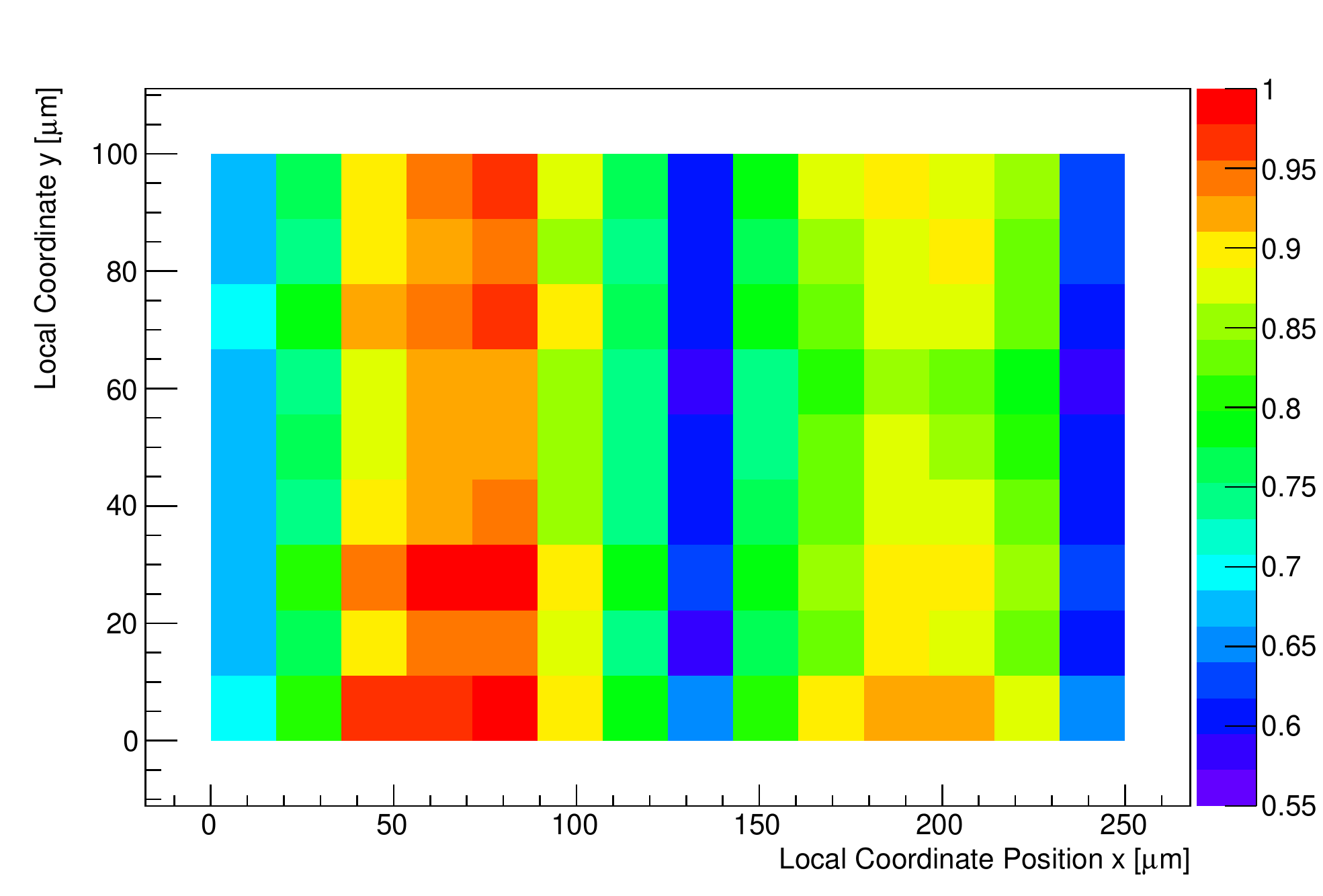}
	\label{fig:sps_c19_ineffmap}
      }
      \subfigure[]{
	\raisebox{5mm}{\includegraphics[width=0.47\textwidth]{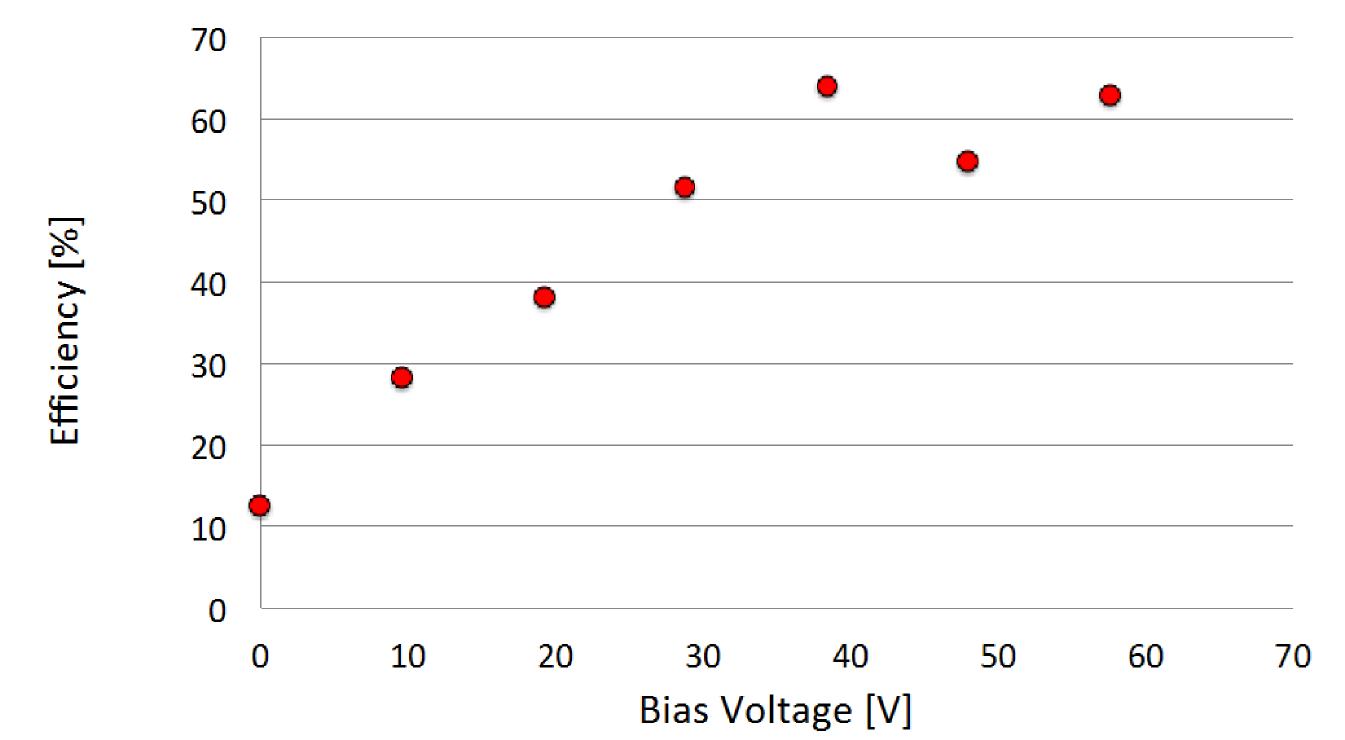}}
	\label{fig:desy_biasscan}
      }
      \vspace{-6pt}
      \caption[]{\subref{fig:sps_c19_ineffmap} Efficiency map of all pixels of C19 overlayed showing intra-pixel inhomogeneities. \subref{fig:desy_biasscan} Detection efficiency versus bias voltage for C07 measured at the DESY testbeam.} 
    \end{center}
  \end{figure}

A list of measured samples can be found in table \ref{tab:samples}.\\
During the testbeam at DESY in 2013 the irradiated version 1 sample C07 was measured and a HV bias voltage scan performed.
The detection efficiency shows an increasing trend with the bias voltage (see figure \ref{fig:desy_biasscan}) as expected.
However, at that time the sample could not be tuned\footnote{The tuning algorithms were not fully implemented in the control software}, so the threshold had to be set very high, resulting in a decreased overall efficiency of $\approx 65\%$.\\
Fully implemented tuning algorithms and a significant reduction of noise induced effects were achieved at the CERN testbeams for measurements with the unirradiated version 2 sample C19.
Each pixel was tuned to a threshold right above the noise edge, resulting in a MPV of the threshold distribution of $\approx 500\ldots600$e (electrons) and a fairly homogeneous detection efficiency of $97\%$ (see figure \ref{fig:ps_c19_eff}).

\begin{figure}[!hb]
    \begin{center}
      \subfigure[]{
	\includegraphics[width=0.47\textwidth]{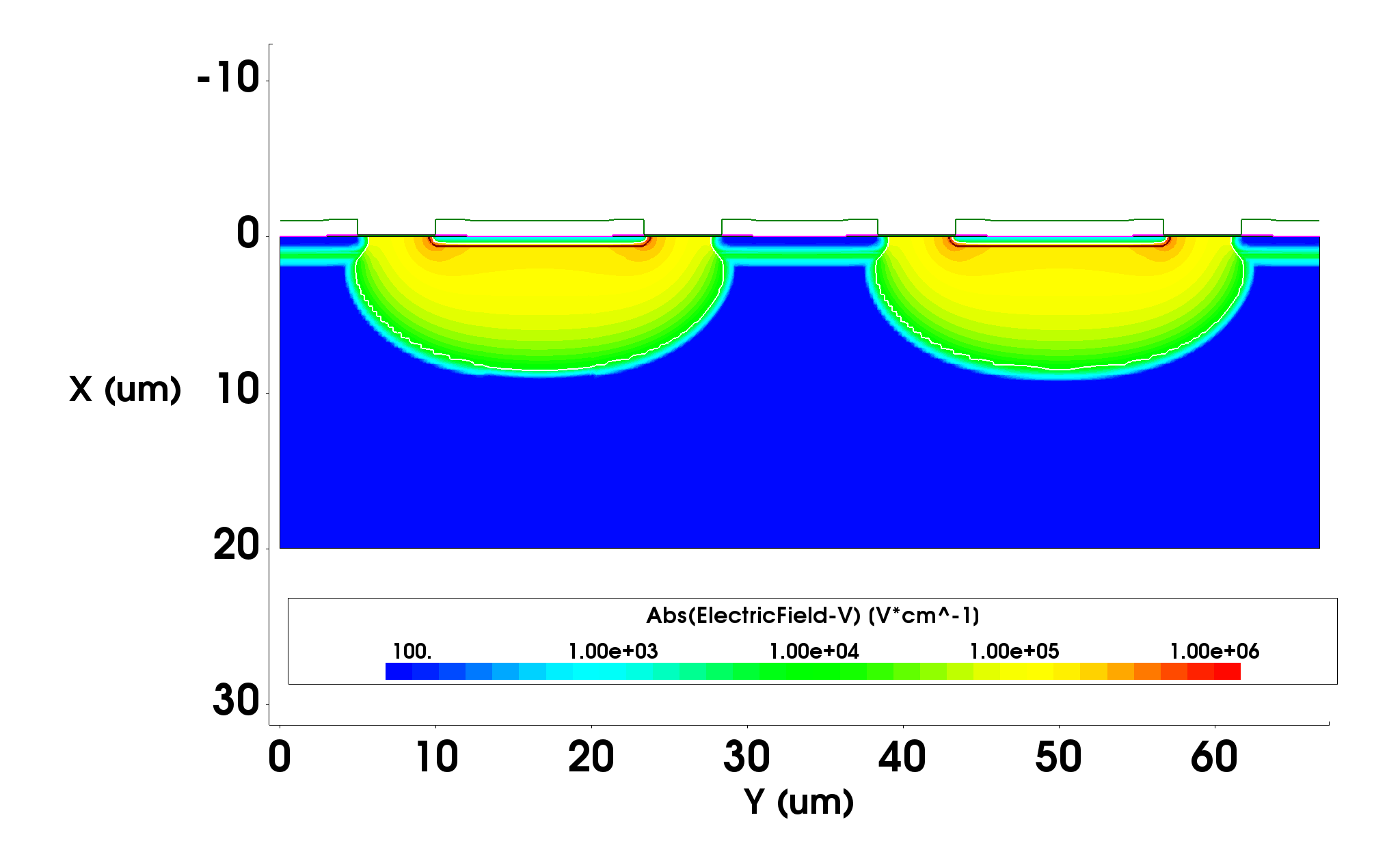}
	\label{fig:tcad_v2}
      }
      \subfigure[]{
	\includegraphics[width=0.47\textwidth]{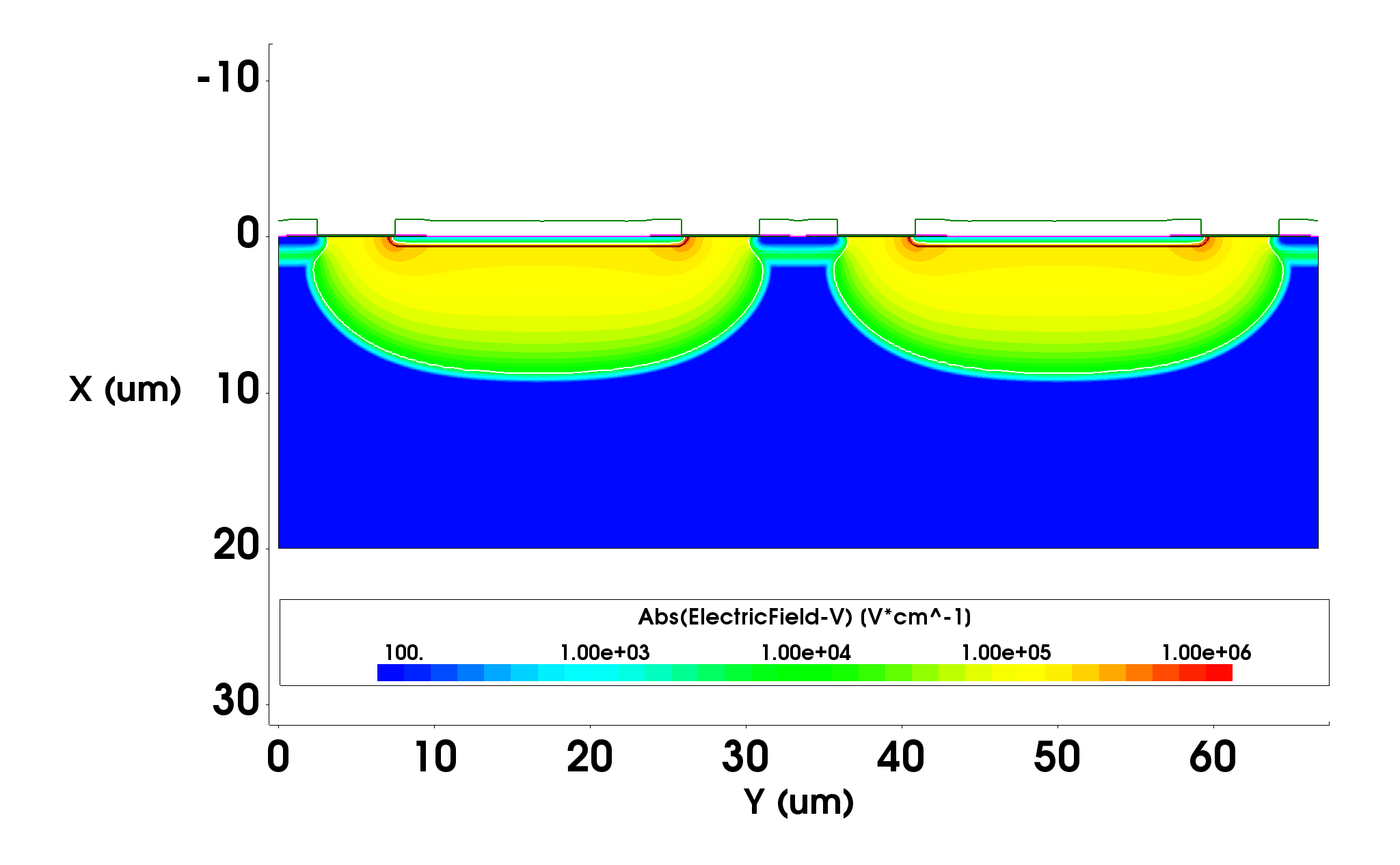}
	\label{fig:tcad_v3}
      }
      \vspace{-6pt}
      \caption[]{Preliminary simulations of the electric field distributions at $U_{Bias}=-80\,$V for the interpixel region of \subref{fig:tcad_v2} HV2FEI4 version 2 and \subref{fig:tcad_v3} version 3 like pixel structures.\cite{mathieu}}
      \label{fig:tcad}
    \end{center}
  \end{figure}

\newpage
The irradiated version 2 sample C22 was measured at a later testbeam at the SPS and showed a mean efficiency of the matrix of ``standard'' pixels of above $95\%$ at a bias voltage of $-80\,$V (figure \ref{fig:sps_c22_eff}).
The pixels were again tuned close to the noise edge to a threshold with an MPV of $\approx 500\,$e.
The lower efficient pixel rows at the bottom end of the plots are of another pixel type that intrinsically has a higher threshold and was not considered in the tuning of the samples.
In order to assess the origin of the missing efficiency of the sensors, dedicated runs with the unirradiated sensor at high thresholds of above $2000$e were taken.
The detection efficiency was resolved with a sub pixel precision and depicted in figure \ref{fig:sps_c19_ineffmap}.
Inefficient regions can be clearly identified between the two sub pixel rows where a ring electrode is implemented distributing the high voltage to the bulk.
Preliminary simulations (see figure \ref{fig:tcad}) suggest that the electrode creates a region of low electric field leading to a reduced charge collection and thus hit detection efficiency.
More narrow electrodes have already been implemented for subsequent versions of the chip and will be tested soon.
  
  \begin{figure}[!hb]
    \begin{center}
      \subfigure[]{
	\includegraphics[width=0.47\textwidth]{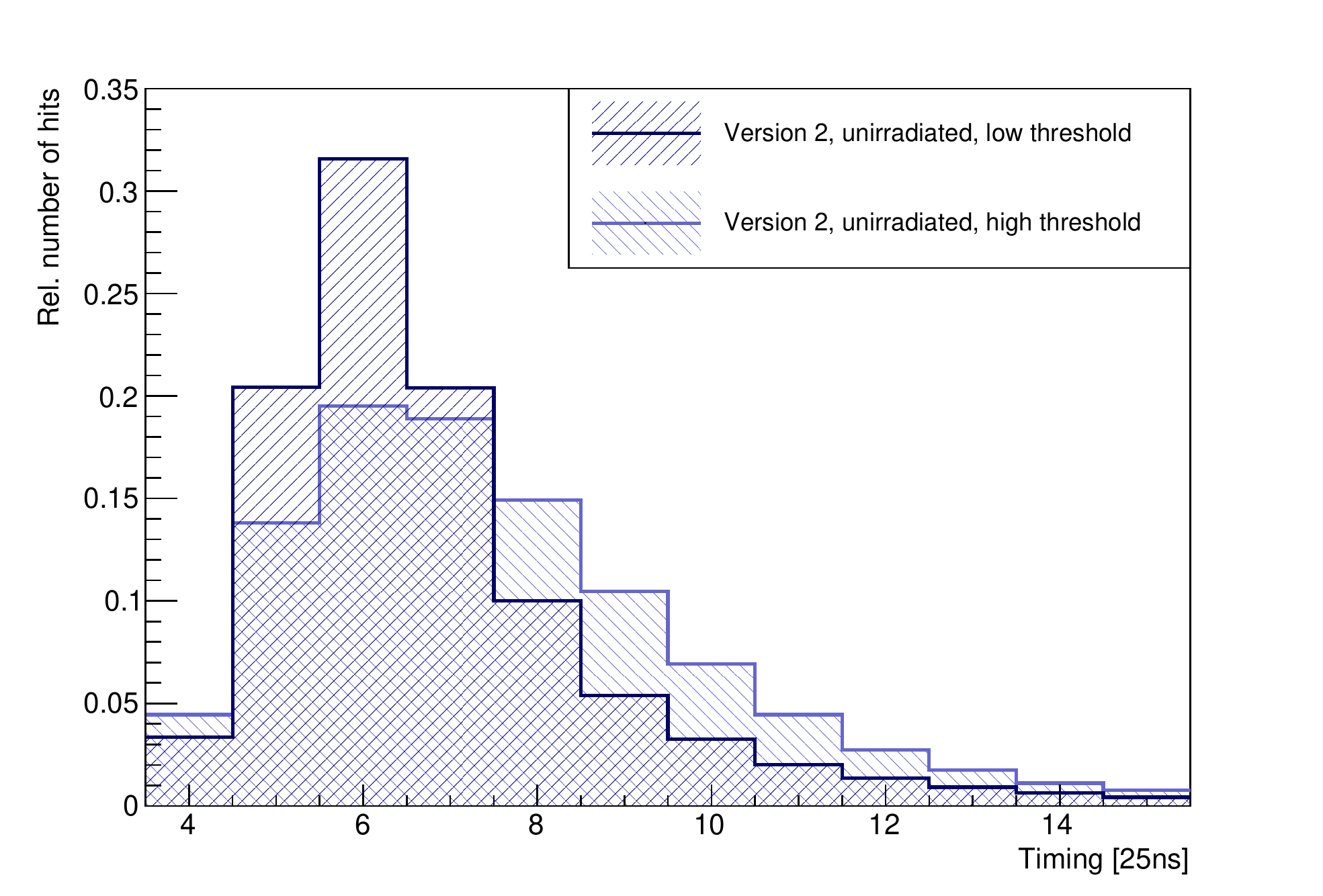}
	\label{fig:c19_lv1}
      }
      \subfigure[]{
	\includegraphics[width=0.47\textwidth]{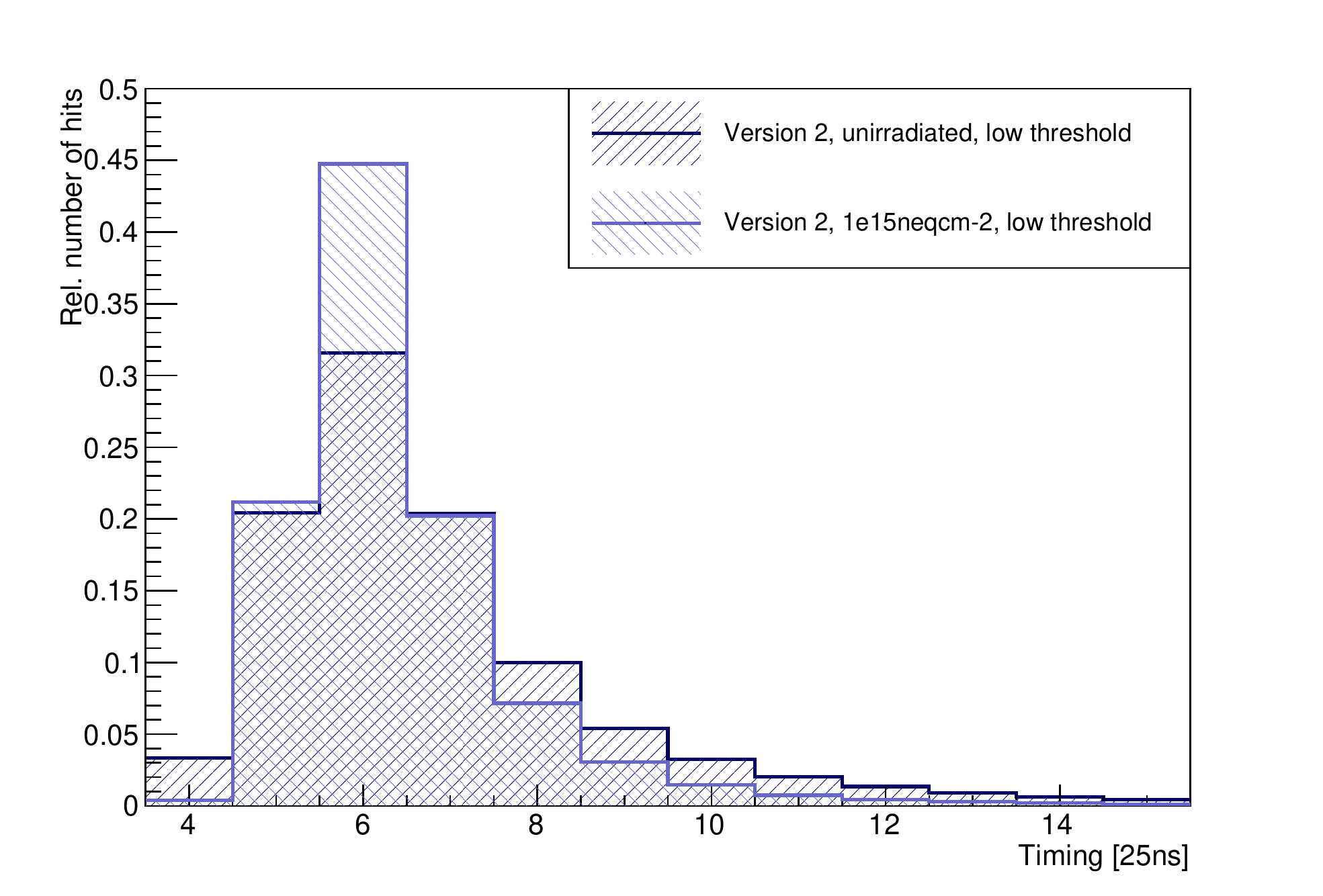}
	\label{fig:lv1_comp}
      }
      \vspace{-6pt}
      \caption[]{(a) Hit timing for \subref{fig:c19_lv1} C19 with low and high threshold settings. \subref{fig:lv1_comp} Comparison of hit timings between C19 and C22.}
      \label{fig:lv1}
    \end{center}
  \end{figure}
During all testbeam measurements an unusually low time resolution of the sensors was observed.
Figure \ref{fig:lv1} shows the hit timing with respect to the telescope trigger in bins of $25\,$ns.
For these plots the noise baseline was subtracted and the resulted histogram normalized to an integral value of 1 in order to allow comparability.
A typical distribution for an unirradiated planar pixel sensor would have entries in maximum two bins.
Figure \ref{fig:c19_lv1} shows the timing of the unirradiated sensor with different thresholds, while figure \ref{fig:lv1_comp} compares the distributions of the unirradiated and the \irrad{1}{15} irradiated sensor.
Although the HV2FEI4 is designed as a drift based sensor, a diffusion component of the generated signal is also contributing, in particular in low field regions (see Edge TCT results).
Depending on the threshold settings the particle detection relies to some extent on the slow diffusion signal as shown in figure \ref{fig:c19_lv1}.
For an \irrad{1}{15} irradiated sensor the diffusion component should be highly suppressed due to trapping.
The timing distributions obtained for such a sensor (see figure \ref{fig:lv1_comp}) support this assumption and also imply that for a low threshold the hit detection is based mainly on the drift component.
Current measurements try to assess if the remaining distribution width can also be understood as a result of a slow amplifier response, which then should be taken into consideration for the next prototype versions.
\section{Edge TCT results}\label{seq:etct}
The Transient Current Technique (TCT) is a commonly used tool for characterization of silicon sensors. 
A typically red or infrared laser beam induces an electrical signal by ionization of the sensor bulk.
Strongly focused laser beams allow for high spatial resolution while a time resolved analogue readout of the resulting current yields information about charge movement in the bulk.
In most cases the laser is shot from the top or bottom of the structure under investigation.
In Edge TCT measurements the laser is shot into a side of the sensor allowing to define the charge deposition depth with high precision \cite{kramberger}.
  \begin{figure}[!ht]
    \begin{center}
      \subfigure[]{
	\raisebox{10mm}{\includegraphics[width=.5\textwidth]{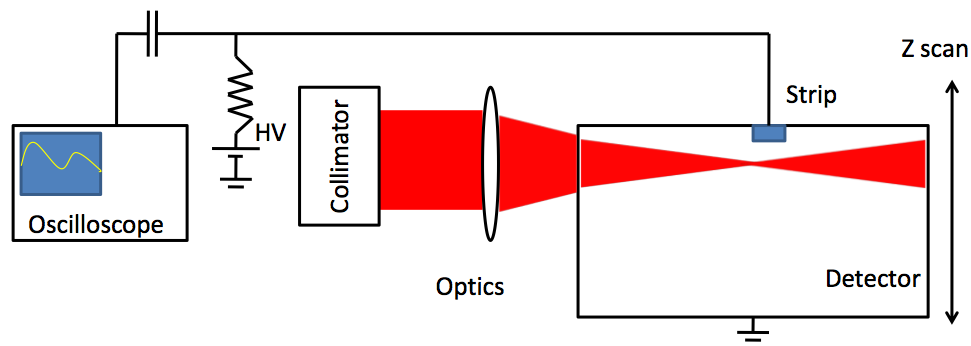}}
	\label{fig:etct-setup}
      }
      \subfigure[]{
	\includegraphics[width=0.45\textwidth]{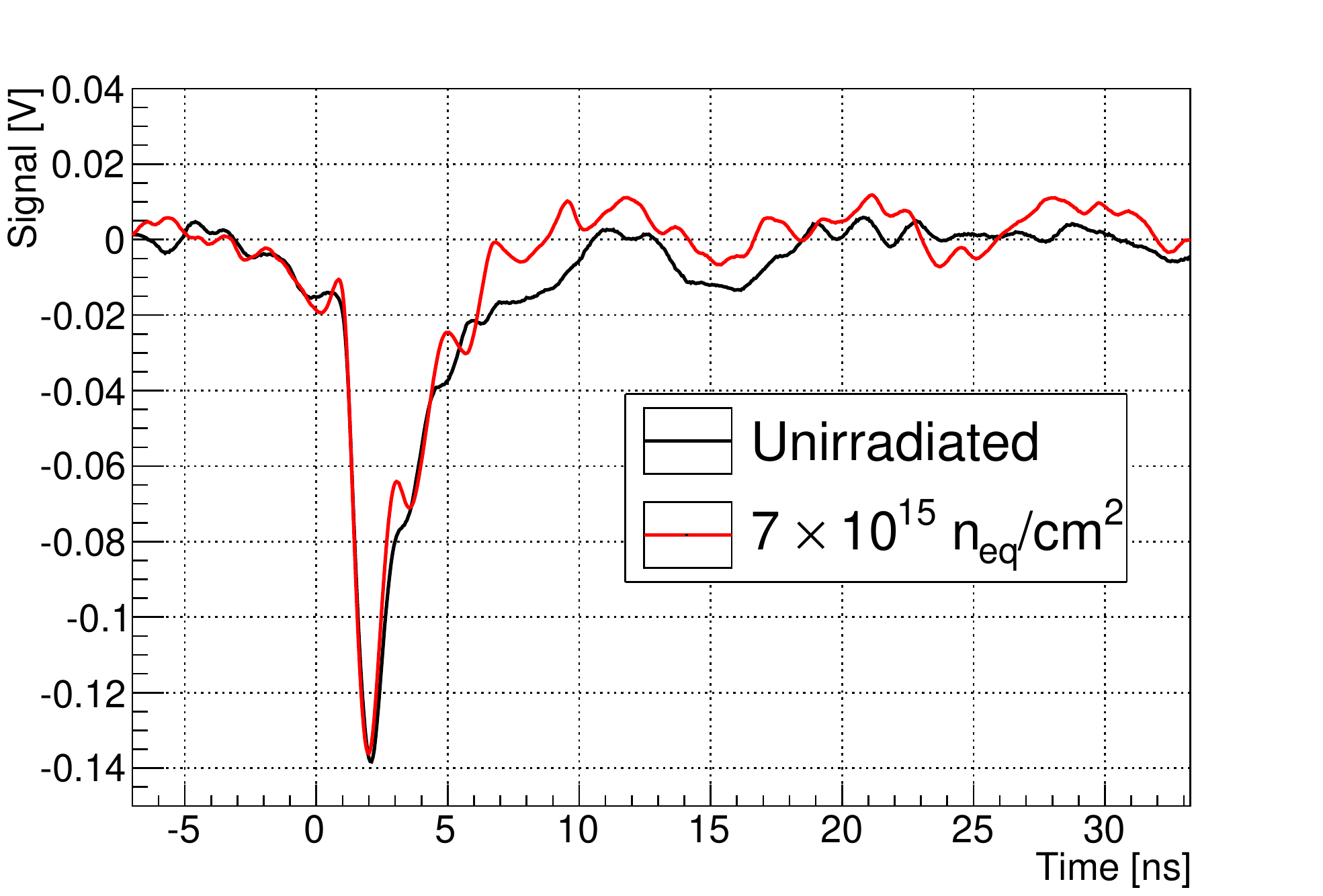}
	\label{fig:etct_signals}
      }
      \vspace{-6pt}
      \caption[]{\subref{fig:etct-setup} Sketch of an Edge TCT setup. \subref{fig:etct_signals} Drift part of signals obtained from the directly accessible deep n-well of an unirradiated and a \irrad{7}{15} irradiated HV2FEI4 version 3 sensor\cite{etct}.} 
    \end{center}
  \end{figure}

\begin{figure}[t!]
    \begin{center}
      \subfigure[]{
	\includegraphics[width=0.47\textwidth]{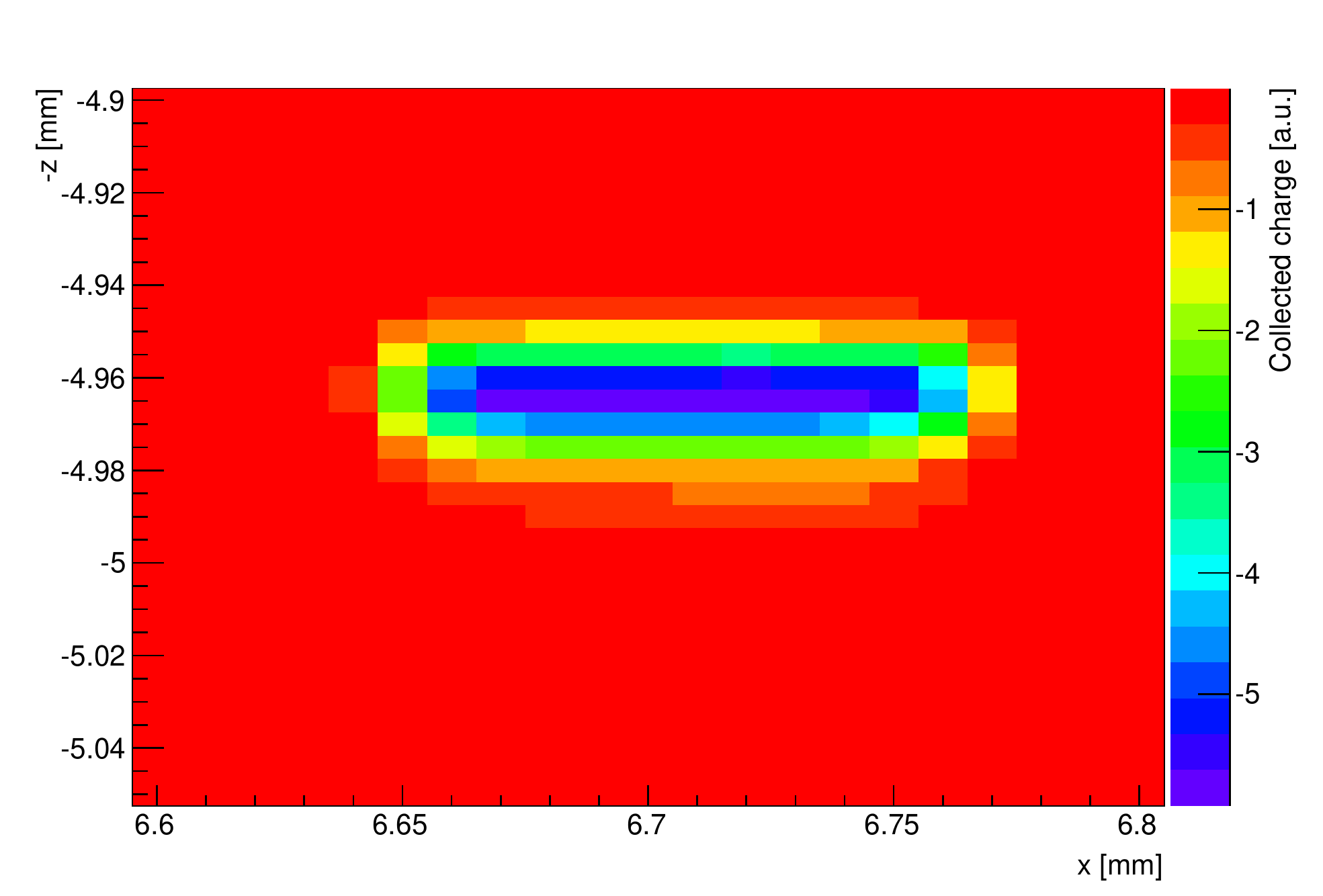}
	\label{fig:unirrad_map_drift}
      }
      \subfigure[]{
	\includegraphics[width=0.48\textwidth]{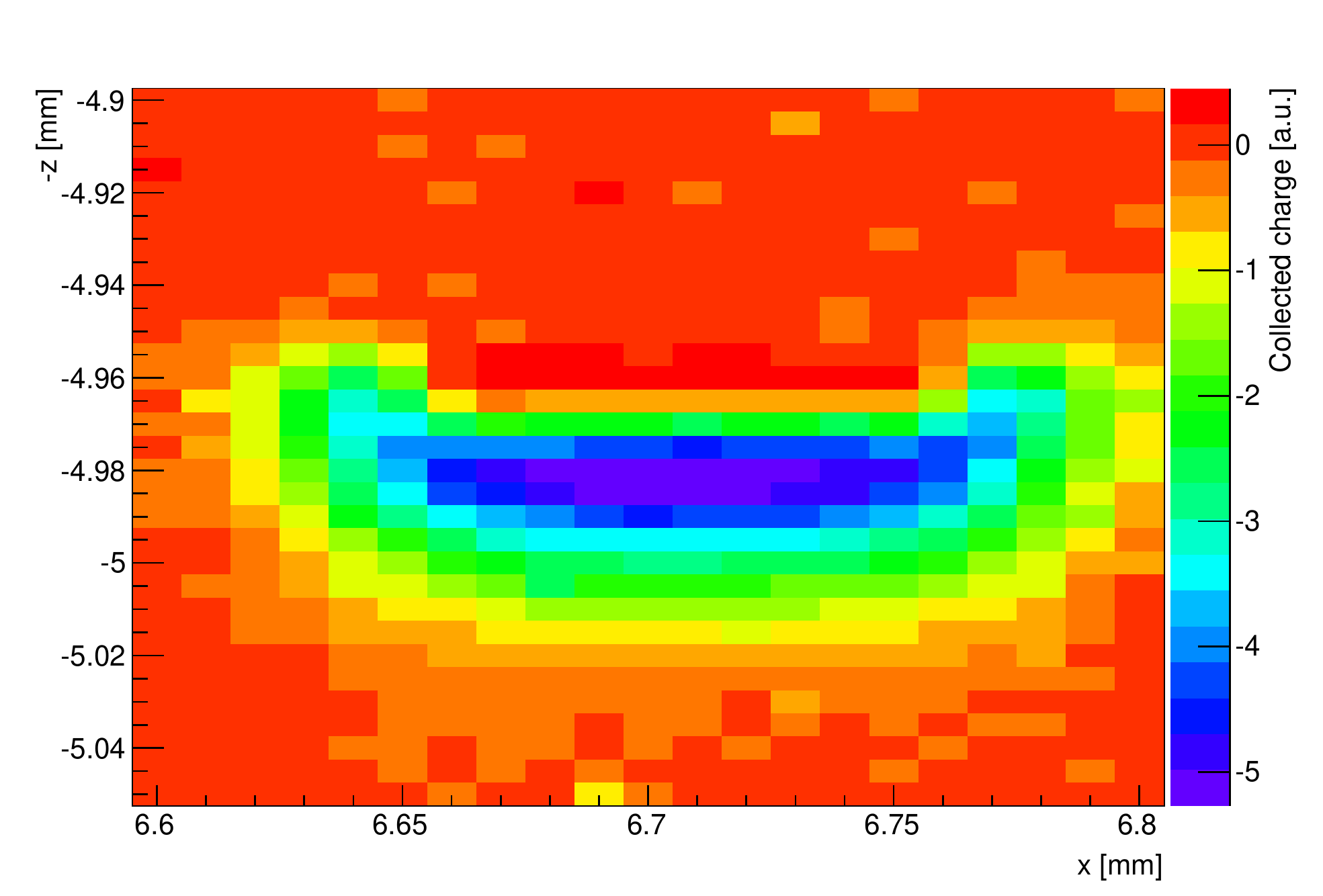}
	\label{fig:unirrad_map_diff}
      }
      \vspace{-6pt}
      \caption[]{Charge collected by the unirradiated sensor originating from \subref{fig:unirrad_map_drift} the drift and \subref{fig:unirrad_map_diff} the diffusion signal at a bias voltage of $-60\,$V\cite{etct}.} 
      \label{fig:unirrad_map}
    \end{center}
  \end{figure}

\begin{figure}[!ht]
    \begin{center}
      \subfigure[]{
	\includegraphics[width=0.45\textwidth]{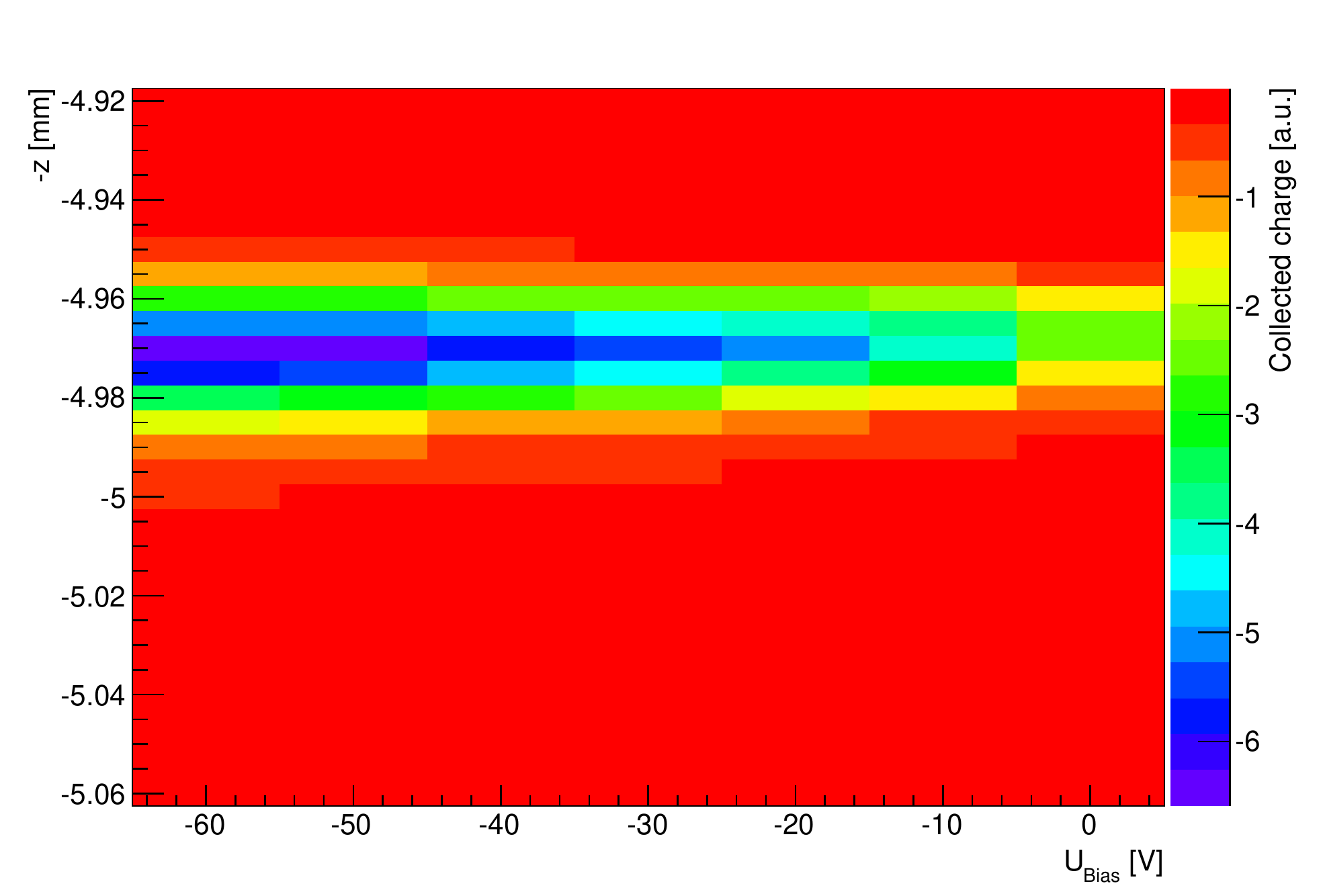}
	\label{fig:unirrad_map_drift_bias}
      }
      \subfigure[]{
	\includegraphics[width=.495\textwidth]{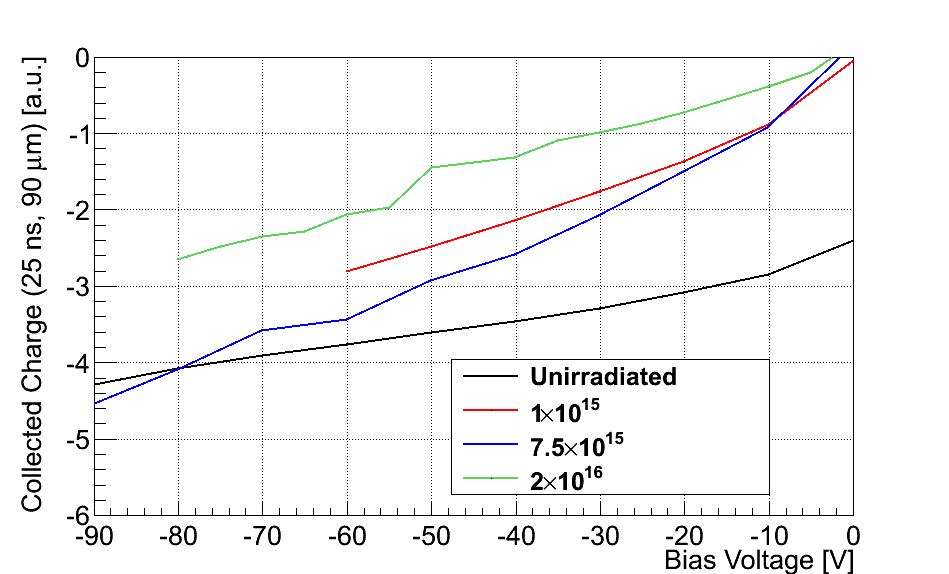}
	\label{fig:etct_comparison}
      }
      \vspace{-6pt}
      \caption[]{\subref{fig:unirrad_map_drift_bias} Depth resolved signal amplitudes for different bias voltages of the unirradiated sensor originating from the drift signal. \subref{fig:etct_comparison} Comparison of charge collected in $25\,$ns by sensors irradiated up to \irrad{2}{16}\cite{etct}.} 
      \label{fig:unirrad_map_bias}
    \end{center}
  \end{figure}

Measurements have been conducted on unirradiated and irradiated HV2FEI4 version 3 samples\cite{etct} with the setup sketched in figure \ref{fig:etct-setup}.
Here a direct contact to the deep n-well is possible, which eliminates the influence of the implemented electronics like in the fully integrated pixel cells.
Lateral scans have been performed to map the charge collection under the deep n-well as well as HV bias scans.
It was observed that the drift signal decays after $2.5\,\text{ns}$ while the slower diffusion signal is spreading over more than $35\,$ns.
Therefore the signals were categorized by applying a timing cut on the recorded waveform.
By plotting the signal amplitudes of an unirradiated sensor for the slow and fast signals with respect to the laser position in figure \ref{fig:unirrad_map} one obtains the position and shape of the depletion zone.\\
The sensor biased to $-60\,$V yields a $\sim 100\,\mu$m wide drift zone with a thickness of $15\,\mu$m which is in agreement with the expectations from the resistivity of the bulk material.
A bias scan, performed in steps of $10\,$V down to $-60\,$V shows a growth of the depletion zone into the bulk (see figure \ref{fig:unirrad_map_drift_bias}).
Additionally, sensors irradiated to \irrad{1}{15}, \irrad{7}{15} and \irrad{2}{16} were measured. 
The unirradiated and \irrad{1}{15} irradiated sensors were operated at room temperature while the others were cooled down to $0^\circ$C.
Waveforms obtained for the unirradiated and the \irrad{7}{15} irradiated sensor at $U_{Bias}=-90\,$V are depicted in figure \ref{fig:etct_signals} and indicate that by applying sufficient bias voltages for fluences of up to \irrad{7}{15} no significant degradation of the drift signal occurs.
Recent measurements of a Sr-90 spectrum, where the collected charge was estimated by integrating over the recorded waveforms suggest a signal degradation of up to $31\,\%$ for \irrad{1}{15} irradiated sensors \cite{simon}.
This can be accounted to the trapping of the slowly diffusing charge outside the drift region.
Compared with the detection timing obtained by testbeam experiments (see section \ref{seq:tbresults}) the slow signal generated by diffusion seems to account for most of the late hits in these measurements.\\
Figure \ref{fig:etct_comparison} shows the signal of different samples for different voltages.
The sensor was scanned along the z-axis and the charge estimated as the integral over $25\,$ns of the signal. 
The sum of the charge per depth is plotted against the bias voltage.
At no applied bias voltage the signal is dominated by diffusion which is highly suppressed for the irradiated sensors.
This is also supported by the absence of degradation of the drift signal due to irradiation as explained above.
For rising bias voltage the charge signal of the irradiated sensors approach the level of the unirradiated one, leading to equal charge collection for the \irrad{7}{15} one at $V_{Bias}=-80\,$V.
At these voltages the charge collection can be considered as purely drift driven.

\newpage

\section{Summary}
Sensors based on AMS H18 High Voltage CMOS technology show promising performance as candidates for the Phase-II upgrade of the ATLAS Inner Tracker.
Production processes are standardized, thus generally cheaper, while simpler interconnection techniques allow for replacement of the costly bump bonding process.
The prototypes can be operated stably after irradiation to ionizing doses and fluences that are expected for the upgraded pixel detector.
Testbeam measurements showed a detection efficiency of above $95\%$ and identified design based inefficient regions in the pixel cells that could be avoided in further submissions, as shown by preliminary simulations.
Edge TCT measurements revealed a strong drift signal independently of the irradiation with the overall signal degradation coming from the trapped diffusion part after irradiation.
For high bias voltages \irrad{7}{15} irradiated sensors demonstrated a charge collection performance comparable to unirradiated sensors.
This will be further investigated by standard TCT and testbeam measurements.
Further lab characterizations and testbeam experiments are conducted on new HV2FEI4 version 4 sensors that implement new charge sensitive amplifiers and an updated position encoding scheme.

\acknowledgments
The work presented in this paper is the result of a highly collaborative endeavor.
We would like to especially thank Christian Gallrapp, Marcos Fernandez Garcia and Michael Moll for the opportunity and the help during the edge TCT measurements.
Further thanks go to Martin Kocian for his constant help in preparing and conducting testbeam measurements and Matevz Cevz and Garrin McGoldrick for establishing and supporting the testbeam reconstruction framework.

The research leading to these results has received funding from the European Commission under the FP7 Research Infrastructures project AIDA grant agreement no. 262025.


\end{document}